\documentclass[aps,prl,twocolumn,superscriptaddress,floatfix,showpacs]{revtex4-1}

\usepackage{amsmath}
\usepackage{amssymb}
\usepackage{graphicx,graphics}
\usepackage{latexsym,verbatim}
\usepackage{dcolumn} 
\usepackage{color}
\usepackage{cancel}
\usepackage{ulem}

\begin{document}

\title{Interactions in Electronic Mach-Zehnder Interferometers with Copropagating Edge Channels}
\author{Luca Chirolli}
 \email{luca.chirolli@icmm.csic.es}
 \affiliation{Instituto de Ciencia de Materiales de Madrid, CSIC, ES-28049 Madrid, Spain}

\author{Fabio Taddei}
\affiliation{NEST, Scuola Normale Superiore and Istituto Nanoscienze-CNR, I-56127 Pisa, Italy}
\author{Rosario Fazio}
\affiliation{NEST, Scuola Normale Superiore and Istituto Nanoscienze-CNR, I-56127 Pisa, Italy}
\author{Vittorio Giovannetti}
\affiliation{NEST, Scuola Normale Superiore and Istituto Nanoscienze-CNR, I-56127 Pisa, Italy}

\begin{abstract}
We study Coulomb interactions in the finite bias response of Mach-Zehnder interferometers,  which exploit copropagating 
edge states in the integer quantum Hall effect. Here, interactions are particularly important since the coherent coupling of 
edge channels is due to a resonant mechanism that is spoiled by inelastic processes. We find that interactions yield a 
saturation, as a function of bias voltage, of the period-averaged interferometer current which gives rise to unusual features, 
such as negative differential conductance, enhancement of the visibility of the current, and nonbounded or even diverging 
visibility of the differential conductance.
\end{abstract}


\maketitle

{\it Introduction. --- } Topological edge states in the integer quantum Hall effect~\cite{QHeffect} represent an ideal 
playground for testing the coherence of electronic systems at a fundamental level.  The harnessing of edge states as quasi-one-dimensional (1D) chiral electronic waveguides has allowed the successful realization of a number of electronic 
interferometric setups, such as those of Mach-Zehnder  \cite{Ji2003,Neder2006,Roulleau2007,Litvin2008} and Fabry-Perot  
\cite{McClure09}. These devices exploit counterpropagating edge states localized at opposite sides of a Hall bar, which are 
brought in contact and mixed by quantum point constrictions (QPCs) that mimic the effect of optical beam splitters (BSs). In 
Mach-Zehnder interferometer (MZI) setups the chirality of electron propagation makes  necessary the adoption of non simply-connected,  Corbino-like geometries, which limit the flexibility of the devices.  In these systems electron-electron 
(e-e) interactions are in general responsible for dephasing via the coupling with external edge 
channels~\cite{ChalkerGefen2007,Marquardt2009}, which manifests as a reduction of the visibility of the Aharonov-Bohm 
(AB) oscillations as a function of the bias voltage.  In particular, puzzling behaviors have been reported 
\cite{Neder2006,Roulleau2007,Litvin2008} in the finite bias response of MZIs, in which the visibility presents a lobe-like 
structure and phase-rigidity \cite{ChalkerGefen2007,Sukhorukov2007,Neder2008,Youn2008,Levkivskyi2008,Sukhorukov2009,Kovrizhin2010,Rufino2012}.

\begin{figure}[t]
\begin{center}
\includegraphics[width=0.45\textwidth]{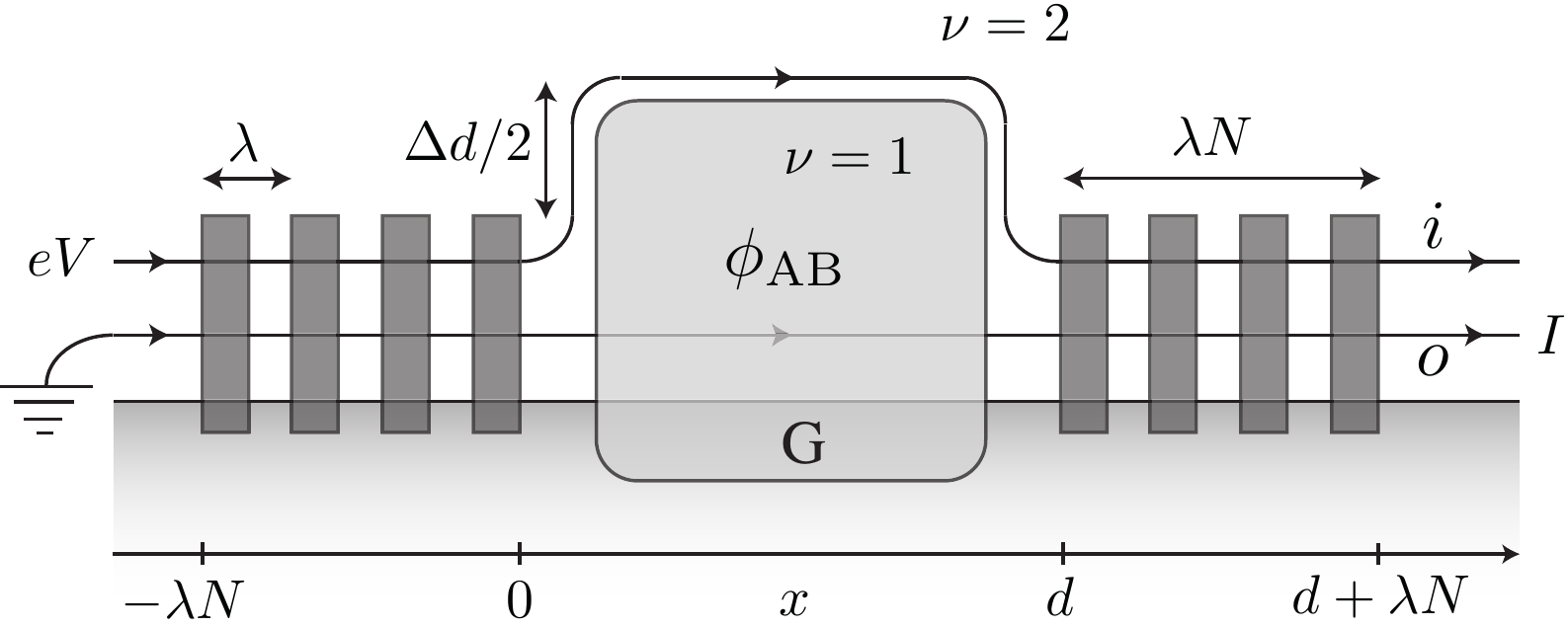}
\caption{Sketch of the MZI. Two sets of $N$ top gates arranged in arrays with periodicity $\lambda$ and separated by the 
distance $d$  represent the $L$ and $R$ BSs of the interferometer. A central top gate  G  locally lowers the filling factor to 
$\nu=1$ and separates the two edge states ($i$ and $o$), which experience a path length difference $\Delta d$ and 
acquire an AB phase $\phi_{\rm AB}$.\label{Fig1}}
\end{center}
\end{figure}

An alternative MZI scheme inducing coherent mixing between edge states copropagating at the {\it same} boundary of 
the sample, has been  suggested~\cite{Giovannetti2008} as a more flexible architecture, which allows multiple device 
concatenation \cite{Giovannetti2008,Chirolli2010} and represents an ideal candidate for implementation of dual-rail 
quantum computation schemes \cite{ChuangYamamoto95,KLM2001}.  As schematized in Fig.~\ref{Fig1},  in such a setup 
the BS transformations are now  implemented  through series of top gates, organized in arrays  of periodicity $\lambda$ 
tuned to  compensate for the difference between the momenta  of the copropagating channels (the inner channel $i$ and 
the outer channel $o$ of the figure) -- see Ref.~\cite{Karmakar2011}, where the first experimental realization of such BSs is reported, and Ref.~\cite{Chirolli2012}.
The corresponding AB phase difference is introduced instead by separating the edge 
modes in the region between the two BSs  through the action of a central top gate \cite{Giovannetti2008}. 
A similar interferometer, but featuring no modulation, has been realized in~\cite{Deviatov}.
The effects  of e-e interactions in these  MZIs are likely to be 
qualitatively different from those observed in the Corbino-like settings of 
Refs.~\cite{Ji2003,Neder2006,Roulleau2007,Litvin2008}. Indeed,  while in the latter the direct coupling between the 
channels that enter  the interferometer can be neglected in the regions where they are coherently mixed (the QPCs), 
in the scheme of Fig.~\ref{Fig1} this is no longer possible due to the non-negligible spatial extension of the top gate arrays. 
This implies a  strong interplay between coherent mixing and interactions which might  impair  the MZI response. Aim of the 
present work is to target such interplay.

As detailed in the following, we show that the inter-edge current $I$ measured at the 
output of the setup of Fig.~\ref{Fig1} possesses a strong non-linear dependence in the bias voltage $V$ that, while still 
exhibiting AB oscillations, leads to saturation of the associated mean value  averaged over the AB phases.  Hence, 
the device presents negative  differential conductance with  unbounded visibility.  Such anomalous behavior occurs since 
interactions enable inelastic scattering which spoils, for increasing voltages, the coherence needed for inter-edge coupling 
to occur at the BSs. Furthermore, as long as the interactions between the edge channels  can be neglected  in the region 
between the two BSs, we also observe that for large voltages  the visibility of current gets enhanced with respect to the 
non-interacting case.

{\it Model: ---} Before  discussing the role of  e-e interactions in the device of Fig.~\ref{Fig1}, we find it useful to  briefly review 
the basic properties of the scheme in the non-interacting case. The underlying idea~\cite{Karmakar2011} is to implement BS transformations between two co-propagating channels ($i$ and $o$),  via the action of a pair of  arrays of top gates (see 
Fig.~\ref{Fig1}) which are spatially modulated at periodicity $\lambda$. Following Refs.~\cite{Karmakar2011,Chirolli2012} 
we describe them through potentials of the form $t_{L}(x)=\bar{t}_L\sin^2(\pi x/\lambda)$, for $-\lambda N<x<0$, and
$t_R(x)=\Bar{t}_R\sin^2(\pi (x-d)/\lambda)$, for $d<x<d+\lambda N$ ($N$ being the number of elements of a single array 
while $d$ being the distance between the two sets as measured according to the coordinates of the channel $o$). Introducing  
the difference $\Delta k=k_i-k_o$ between the momenta $k_i$ and $k_o$ of the two edges, the tunneling amplitude at a given 
BS can then be expressed as $\bar{t}_{\alpha}{\rm sinc}(\lambda N(\Delta k-2\pi/\lambda)/2)$,  at lowest orders in 
$\bar{t}_{\alpha}$ (here $\alpha=$ L,R indicates  the left and right BS, respectively)~\cite{Karmakar2011}. In this scenario 
Mach-Zhender interferences can be observed  by introducing between the two BSs  a top gate which locally lowers the filling 
factor to $\nu=1$ and diverts the inner edge state toward the interior of the mesa \cite{Giovannetti2008}. This way, the two 
channels are guided along paths of difference lengths, $d_o=d$ and $d_i=d+\Delta d$, thus acquiring an AB phase 
difference $\phi_{\rm AB}$ proportional to the magnetic field $B$ and to the area enclosed by the path and a dynamical 
phase. The transmission probability $T(\epsilon)$ at energy $\epsilon$ of the MZI, from inner channel on the left to outer 
channel on the right, is then given by 
$T(\epsilon)\propto [|\bar{t}_L|^2+|\bar{t}_R|^2+2|\bar{t}_L\bar{t}_R|\cos(\phi_{\rm AB}+\epsilon\Delta d/v_F)]S^2$, where 
$v_F$ is the group velocity of edge states and $S={\rm sinc}(\lambda N(\Delta k-2\pi/\lambda)/2)$ is the BS amplitude which 
is optimal when the resonant condition $\Delta k= 2\pi/\lambda$ is met. A bias voltage $V$ applied between channels $i$ 
and $o$  gives rise to a zero-temperature current $I(V)\propto \int_0^{eV}d\epsilon~ T(\epsilon)$,  whose associated visibility  
${\cal V}_I=({\rm max}_\phi I-{\rm min}_\phi I)/({\rm max}_\phi I+{\rm min}_\phi I)$ of the AB oscillations amounts to 
${\cal V}_I={\cal V}_\sigma^{(0)}|{\rm sinc}(eV\Delta d/2v_F)|$, with 
${\cal V}_\sigma^{(0)}=2|\bar{t}_L\bar{t}_R|/(|\bar{t}_L|^2+|\bar{t}_R|^2)$ 
being the oscillation visibility of the differential conductance $\sigma=dI/dV$.

To analyze the effect of interactions in such a set up we describe the linearly dispersing electronic excitations around the 
Fermi energy by means of two chiral fermion fields  $e^{ik_mx}\psi_m(x)$, with $m=i,o$, each propagating at mean momentum 
$k_m$. The kinetic Hamiltonian can then be written as ($\hbar=1$) $H_{\rm kin}=-iv_F\sum_m\int dx\psi_m^{\dag}\partial_x \psi_m$. 
The action of the  BSs are instead assigned by means of the tunneling Hamiltonian 
$H_{\rm tun}=\sum_{\alpha=L,R}(A_\alpha+A^{\dag}_{\alpha})$, where $A_\alpha$ describe  the action of the gate arrays 
potentials and are defined as 
$A_L=\int dx~t_L(x) e^{i\Delta k x}\; \psi_o^{\dag}(x)\psi_i(x)$, 
$A_R=\int dx~t_R(x)e^{i\Delta k x} \; \psi_o^{\dag}(x)\psi_i(x+\Delta d)$
with $t_L(x)$, $t_R(x)$, and $\Delta d$ introduced previously. In these expressions the local phase shift $e^{i\Delta k x}$ 
accounts for the resonant behavior of the MZI. The AB phase of the setup (together with a dynamical phase term 
$k_id_i-k_od_o$)  is instead included in the tunneling amplitude of the right beam-splitter, i.e. 
$\bar{t}_R\to e^{i\phi_{\rm AB}}\bar{t}_R$.

As far as  e-e interactions are concerned, an electron propagating in one edge channel can interact with all the electrons 
in the Fermi seas of both channels. Although the precise form of the interaction potential is unknown, screening provided 
by top gates makes it is reasonable to assume a short range density-density interaction. The latter however needs not to 
be uniform in the whole sample: as a matter of fact, while  intra-channel couplings  are present everywhere, the inter-channel 
interactions depend  on the edge channel spatial separation, which in our setup varies strongly from place to place (see 
Fig.~\ref{Fig1}). In particular in the region between the BSs  edge states are brought far apart by the central top gate G and 
it is reasonable to assume the inter-channel coupling  to be off. Inter-channel interactions, however, cannot be excluded in 
the regions before and after G, where  the tunneling term $H_{\rm tun}$ is active. Indicating with  $\rho_m=\psi^{\dag}_m\psi_m$  
the 1D density operator in channel $m=i,o$,  we account for these effects by introducing an inter-channel e-e coupling 
$H_{\rm inter}= \int dx \int dx'  :\rho_o(x) U(x,x')\rho_i(x'):$ characterized by a coordinate dependent potential $U(x,x')$ which 
nullifies in the central top gate region (i.e.  $0\lesssim  x \lesssim  d$) and which approaches the short range  behaviors 
${2 \pi g} \delta(x-x')$  and ${2 \pi g} \delta(x-x'-\Delta d)$ on the lhs part (i.e. $x\lesssim  0$) and on the rhs part ($x\gtrsim d$) 
of the setup, respectively (the transition between these regions being smooth). In these expressions the interaction strength 
$g$ has the dimension of a velocity while, similarly to $H_{\rm tun}$,  the  parameter $\Delta d$ accounts for the relative 
coordinate shift experienced by the inner edge  with respect to the outer. A short range intra-channel coupling term of the 
form  $H_{\rm intra}=\pi u \sum_m\int dx:\rho^2_{m}(x):$ is also considered (in this case however no coordinate dependence 
is  assumed).

{\it Methods: ---}  Including the e-e interaction, the response of the interferometer driven out from equilibrium by a bias voltage 
$\mu_i-\mu_o=eV$ applied between the two edge states, will be analyzed at lowest order in the tunneling Hamiltonian 
$H_{\rm tun}$ (an exact analytical treatment being impossible). At second order in the  amplitudes $\bar{t}_\alpha$ this allows 
to express the current flowing through the outer edge as $I(V)=e\sum_{\alpha,\beta}\int dt\langle[A_\alpha^{\dag}(t),A_\beta(0)]\rangle$, where $A_{\alpha}(t)=e^{iH_0t}A_{\alpha}e^{-iH_0t}$ is the tunneling term evolved through the kinetic and interaction components 
of the system  Hamiltonian, i.e.  $H_0=H_{\rm kin}+H_{\rm intra}+H_{\rm inter}$, while expectation values are taken with respect 
to the ground state of the system biased by the chemical potential difference $eV$~\cite{SukhorukovBurkardLoss,ChalkerGefen2007}.

Expanding the summation over $\alpha$ and $\beta$, we recognize the presence of three contributions:  $I=I_L+I_R+I_\Phi$ 
with $I_\alpha=e\int dt\langle[A_\alpha^{\dag}(t),A_\alpha(0)]\rangle$, for $\alpha=L,R$ being direct terms which do not depend  
upon the relative phase accumulated by the electrons when traveling through the MZI, and with 
$I_\Phi=e\int dt\langle[A^{\dag}_L(t),A_R(0)]\rangle+{\rm c.c.}$ being a  cross-term, which is sensitive to the $AB$ phase. Explicit expressions for these quantities  are obtained by means of the two-point electron and hole Green's functions 
${\cal G}^e_m(x,t;x')=\langle\psi_m(x,t)\psi^{\dag}_m(x',0)\rangle$ and ${\cal G}^h_m(x,t;x')=\langle\psi^{\dag}_m(x,t)\psi_m(x',0)\rangle$,  
which we calculate by bosonization of the Hamiltonian $H_0$. Following the formalism of Refs.~\cite{ChalkerGefen2007,Levkivskyi2008,vonDelft},  we introduce chiral bosonic fields $\phi_m$, which 
satisfy the  commutation rules $[\phi_m(x),\phi_{m'}(x')]=i\pi\delta_{m,m'}{\rm sign}(x-x')$ \cite{vonDelft,Giamarchi} and express the 
fermion fields as $\psi_m= \frac{\hat{F}_m}{\sqrt{2\pi a}} e^{2\pi i \hat{N}_m x/L} e^{-i\phi_m}$ where $a$ is a cutoff length that 
regularizes the theory at short wavelengths,  $\hat{F}_m$ are the Klein operators, $\hat{N}_m=\int dx :\rho_m(x):$ are the total 
number  operators of the  edge states, while finally $L$ is the edge quantization lengths. Observing that 
$\rho_m=(1/2\pi)\partial_x\phi_m + \hat{N}_m/L$,  the kinetic and intra-channel Hamiltonian (apart for an irrelevant term 
proportional to $\sum_m \hat{N}_m$) becomes 
$H_{\rm kin}+H_{\rm int}^{\rm intra}=(v/4\pi)\sum_m\int dx\;(\partial_x\phi_m)^2 + H^{\rm intra}_C$ where $v=v_F+u$ is the 
renormalized edge group velocity~\cite{ChalkerGefen2007,Levkivskyi2008,vonDelft},  and where 
$H^{\rm intra}_C=\pi v \sum_{m} \hat{N}_m^2 /L$ is a capacitive contribution. Vice-versa, exploiting the smooth variation 
assumption of the potential $U(x,x')$ and the fact that $L$ is the largest length in the problem, the inter-channel interaction 
term yields
\begin{eqnarray}\label{Eq:Hinter}
&& H_{\rm int}^{\rm inter}\simeq g\int_{-\infty}^0\frac{dx}{2\pi}(\partial_x\phi_o)(\partial_x\phi_i)\nonumber\\
&&+ g\int^{\infty}_d\frac{dx}{2\pi}(\partial_x\phi_o(x))(\partial_x\phi_i(x+\Delta d))+ H_C^{\rm inter},
\end{eqnarray}
with  $H_C^{\rm inter}= (2\pi g/L)\hat{N}_o\hat{N}_i$ being a cross capacitive contribution. In contrast with Ref.~\cite{Levkivskyi2008}, the inter-channel interaction Eq.~(\ref{Eq:Hinter})  is specifically active only in the tunneling regions. This will lead to qualitatively different results. The Hamiltonian $H_0$ can now be brought 
to a diagonal form by solving the eigenvector equation $[H_0,\hat\gamma_\pm(\epsilon)]+\epsilon \;\hat\gamma_\pm(\epsilon)=0$, 
which defines bosonic energy eigenmodes $\hat\gamma_\pm(\epsilon)$. Accordingly we obtain 
\begin{equation}
\phi_m(x,t)=\int_0^{\infty}\frac{d\epsilon}{\sqrt \epsilon}e^{-i\epsilon (t-i\tau)}
\sum_{s=\pm}\varphi^s_m(x,\epsilon)\hat\gamma_s(\epsilon)+{\rm h.c.}, \nonumber 
\end{equation}
where the wavefuctions $\varphi^s_m$ satisfy the relation $\sum_s\int_0^{\infty}\frac{d\epsilon}{\epsilon}{\rm Im}
[\varphi^s_m(x,\epsilon)\varphi^s_{m'}(x',\epsilon)^*]=\frac{\pi}{2} \delta_{m,m'}{\rm sign}(x-x')$, and $1/\tau>0$ is an energy 
cutoff  related to $a$  via the non-interacting dispersion $\epsilon=v_Fq$, i.e.  $a/\tau=v_F$. Solving the equations of motion by 
requiring only continuity of the wave functions we find $\varphi_m^\pm=f_\pm(x)$ for $x<0$, $\varphi_m^\pm=C_\pm e^{i\epsilon x/v}$ 
for $0<x<d_m$, with $C_+=1$, $C_-=0$, and $\varphi_m^\pm=e^{i\epsilon d_m/v}f_\pm(x-d_m)$ for $x>d_m$, in terms of the 
symmetric and antisymmetric combinations $f_\pm(x)=(e^{i\epsilon x/v_+}\pm e^{i\epsilon x/v_-})/2$. Two new velocities enter the 
problem, a fast charged mode which propagates at $v_+=v+g$ and a slow neutral mode which propagates at $v_-=v-g$. Exploiting 
these results, the electron and hole Green's functions can be finally written as ${\cal G}_m^{e}=e^{-i\mu_m(t-(x-x')/v_-)}{\cal G}_m(x,t;x')$ and ${\cal G}_m^{h}=e^{i\mu_m(t-(x-x')/v_-)}{\cal G}_m(x,t;x')$, with the zero-bias Green's function 
${\cal G}_m(x,t;x')=\frac{1}{2\pi a}\langle e^{i\phi_m(x,t)}e^{-i\phi_m(x',0)}\rangle$, which, thanks to the quadratic nature of the 
bosonized Hamiltonian $H_0$, can be easily computed  in terms of the wave functions $\varphi^s_m$.  In particular 
to compute the direct current terms $I_L$ and $I_R$ we only need the correlation functions in the BS  regions: $x,x'<0$ for the 
left BS, and $x,x'>d_m$ for the right BS.  At zero temperature, for these combinations we find ${\cal G}_m=\tfrac{i}{2\pi v_F}X_+^{-1/2}X_-^{-1/2}$, with $X_s=(x-x')/v_s-t+i\tau$, in agreement with Ref.~\cite{Levkivskyi2008} (for finite temperatures see~\cite{Note}).  
The  term $I_\Phi$ is obtained instead through crossed combinations. In particular  for $x>d_m$ and $x'<0$ we get 
${\cal G}_m=\frac{i}{2\pi v_F}\prod_{s=\pm}[X_s+d_m(1/v-1/v_s)]^{-1/2}$,
and analogously for $x'>d_m$ and $x<0$, with the replacement $d_m\to -d_m$.

\begin{figure}[t]
\begin{center}
\includegraphics[width=0.45\textwidth]{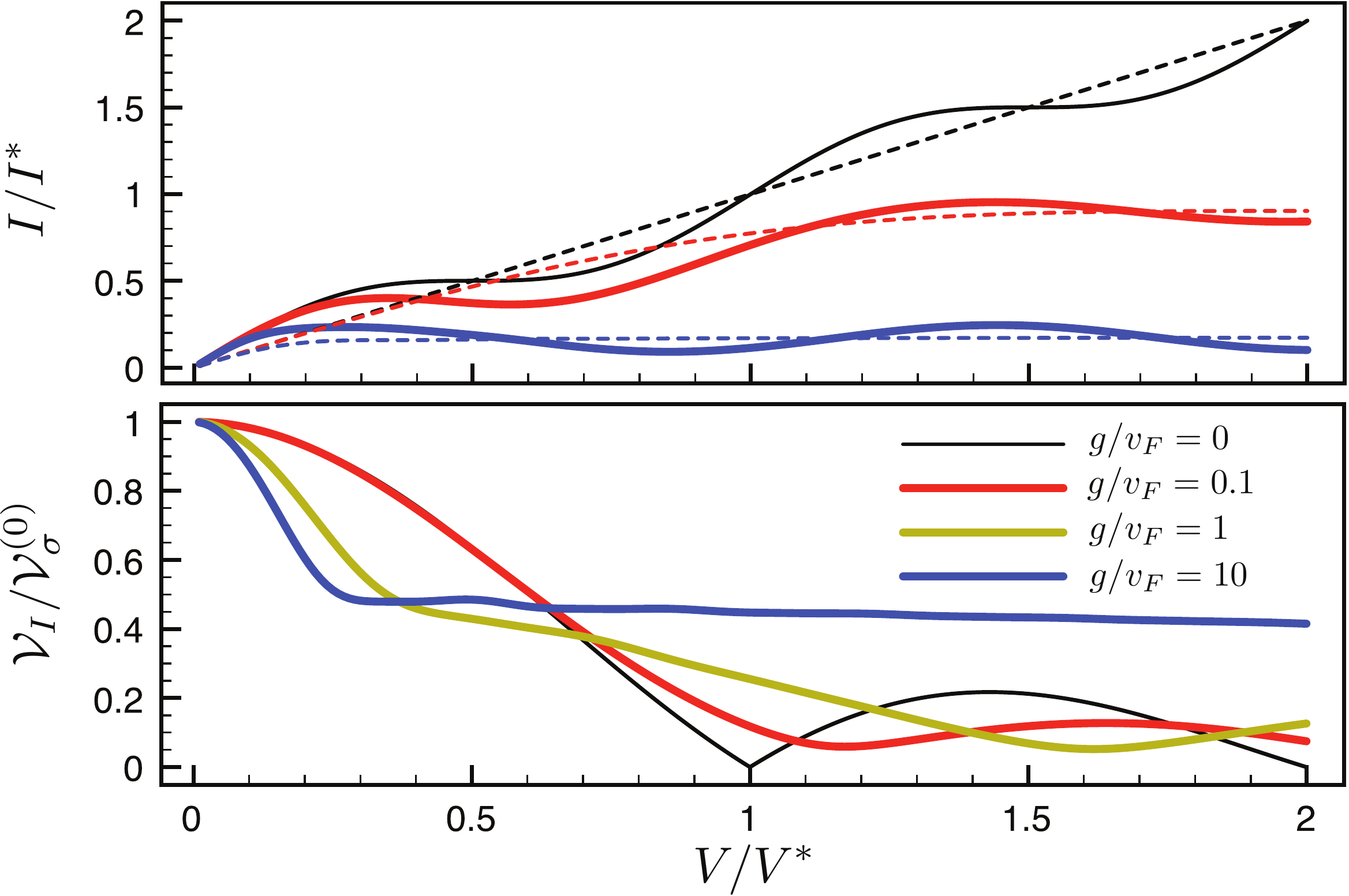}
\caption{(Color online) Response of the MZI versus bias $V$ [in units of $V^*=2\pi v_F/(e \Delta d)$] 
for different values of the interaction parameter $g/v_F$. We have assumed $u=g$, $\phi_{\rm AB}=0$, 
$|\bar{t}_L|=|\bar{t}_R|=|\bar{t}|$ and that the resonant condition, $\Delta k=2\pi/\lambda$, is satisfied. 
Top: total current $I$ (full curves) and BS current (dashed curves) in units of 
$I^*=n_F^2 |\bar{t}|^2e^2V^*/\pi$. Bottom: Visibility of the AB oscillations in the current 
${\cal V}_I$ in units of ${\cal V}_\sigma^{(0)}$.  According to Ref.~\cite{Karmakar2011}, we set 
$\lambda N/\Delta d=3$ and $d/\Delta d=1$.  
\label{Fig2}}
\end{center}
\end{figure}

{\it Currents: ---} With the help of the Green's functions, the  $I$-$V$ characteristics of the setup 
can now be explicitly computed. In particular for the direct contributions of the current one gets 
\begin{eqnarray}\label{InteractingCurrent}
I_\alpha(V)&=&\frac{e}{2\pi}n_F^2|\bar{t}_\alpha|^2\int_0^{eV} d\epsilon S^2(\epsilon/\epsilon_c),
\end{eqnarray}
with the resonance function $S(x)={\rm sinc}(\lambda N(\Delta k-2\pi/\lambda)/2-x/2)$, and the density of states at the Fermi 
energy $n_F=\lambda N/(2v_F)$. The energy scale $\epsilon_c=(\lambda N(1/v_--1/v_+))^{-1}$ is associated to the difference 
in time-of-flight for propagation of the bosonic excitations at speeds $v_+$ and $v_-$ through the BS of length $\lambda N$: in 
the absence of interactions it diverges. We notice that while for a low bias $I_\alpha$ is linear in $V$, for a bias larger than 
$\epsilon_c$ it saturates to the  constant value $I^{\rm asy}_\alpha=\pi G_0n_F^2|\bar{t}_\alpha|^2\epsilon_c$. This is shown 
by the dashed lines in Fig.~\ref{Fig2}, top panel, where $I_\alpha$ is plotted as a function of $V$ for different values of $g$. Such 
behavior is due to the fact that inelastic processes, which are induced by the interaction and increase with increasing voltage bias, 
spoil the resonant coherent effect which is responsible for the transfer of charges between the two edge channels in the BS, 
thus suppressing $I_{\alpha}$.

The cross term contribution $I_\Phi$ to the  current  is obtained instead through integration over the branch 
cuts~\cite{Levkivskyi2008} of the Green's functions in the cross region. Introducing 
\begin{equation}%\label{A} 
{\cal A}=\int_0^{eV}d\epsilon e^{i\varphi_V(\epsilon)}S^2(\epsilon/\epsilon_c)
J_0\left[\frac{\epsilon}{2\epsilon_{\rm dyn}}\right]J_0\left[\frac{eV-\epsilon}{2\epsilon_{\rm dyn}}\right], \nonumber 
\end{equation}
together with the phase $\varphi_V(\epsilon)=\epsilon/\epsilon_c+(\Delta k-2\pi/\lambda)\lambda N+
eV[\Delta d/(2v_-)+(d+\Delta d/2)(1/v-1/v_-)]$,  we obtain
\begin{equation}%\label{Iphi}
I_\Phi(V)=\frac{e}{2\pi}n_F^22|\bar{t}_L\bar{t}_R||{\cal A}|\cos(\phi_{\rm AB}+{\rm arg}({\cal A})), \nonumber 
\end{equation}
where  $J_0[x]$ is the Bessel function of the first kind  and  $\epsilon_{\rm dyn}=v/\Delta d$. The latter is a new energy scale
associated to the dynamical phase difference acquired by the electrons in the interference region between the BSs, where 
inter-channel interaction is absent and excitations move at speed $v$. Due to the bias dependence of ${\rm arg}({\cal A})$, 
the current $I_\Phi$ shows oscillations versus bias around zero.

The overall $I$-$V$ characteristics (full lines plotted for different values of $g$ in Fig.~\ref{Fig2}, top panel) show an oscillating 
behavior which becomes non-monotonic for large values of the bias $V$. The associated visibility is
${\cal V}_I(V)={\cal V}_\sigma^{(0)}|{\cal A}|\left/\int_0^{eV}d\epsilon ~S^2(\epsilon/\epsilon_c)\right.$, where ${\cal V}^ {(0)}_\sigma$, 
as in the absence of interactions, provides the upper bound (${\cal V}_I\leq {\cal V}_\sigma^{(0)}\leq 1$). Plots of ${\cal V}_I(V)$  for 
different values of $g$ are shown in the bottom panel of Fig.~\ref{Fig2}. For $g/v_F=0.1$ (red curve), weakly interacting regime, 
${\cal V}_I(V)$  closely follows the non-interacting case (black curve) for $V\leq V^*=2\pi v_F/(e \Delta d)$ and thereafter 
oscillating without reaching zero.  In the strongly interacting regime $g\gg v_F$, the scale $\epsilon_c$ already dominates at 
relatively low biases, leading to a completely different behavior. As shown by the blue curve ($g/v_F=10$), ${\cal V}_I(V)$ 
decreases with the bias very rapidly up to $eV\leq \epsilon_c$ and very slowly thereafter, on the scale $\epsilon_{\rm dyn}$  ($2\pi \epsilon_c/(eV^*)\simeq 0.32$ and $2\pi \epsilon_{\rm dyn}/(eV^*)\simeq 11$, with the parameters used in Fig.~\ref{Fig2}). The almost constant value exhibited by the curve is the consequence of the fact that the current 
oscillates around a constant value, with amplitude of oscillations that decreases very slowly with bias. Interestingly, in this regime 
the visibility in the interacting case is higher than in the non-interacting case. Very peculiar is also the case of a symmetric 
interferometer, $\Delta d=0$, where $\epsilon_{\rm dyn}$ diverges and the visibility becomes 
${\cal V}^{\Delta d=0}_I={\cal V}_\sigma^{(0)}|{\cal F}_2(V)|/{\cal F}_0(V)$, where 
${\cal F}_p(V)=\int_0^{eV/(2\epsilon_c)}dx~ e^{ixp}~{\rm sinc}^2(x)$. We find that the ratio 
${\cal V}^{\Delta d=0}_I/{\cal V}_\sigma^{(0)}$ goes, for $eV\gg \epsilon_c$, asymptotically to the finite value $2\ln(2)/\pi$. 
Therefore, a symmetric device over-performs with respect to an asymmetric one for large voltages.

The highly non-monotonic behavior of $I(V)$  detailed above  is best understood by studying the differential conductance 
$dI/dV$. In particular, its direct contribution amounts to 
$\tfrac{d I_\alpha}{d V}=\frac{e}{2\pi} n_F^2|\bar{t}_\alpha|^2S^2(eV/\epsilon_c)$, 
and it shows how e-e interactions effectively give rise to an interaction- and bias-dependent shift 
of the resonance condition
\begin{equation}\label{Eq:resonance}
\Delta k=\frac{2\pi}{\lambda}+\frac{2g}{v_+v_-}eV.
\end{equation}
As a result, $dI_\alpha/dV$ becomes negligible beyond an applied voltage on order of $\epsilon_c$, which explains the saturation 
of the period-averaged current in Fig.~\ref{Fig2} in the strongly interacting regime.  At the same time a non-negligible cross term $dI_\Phi/dV$ 
results in an overall behavior of the total differential conductance $dI /dV$ which exhibits regions of negative values. Furthermore, 
while the associated visibility of the AB oscillations in the presence of interactions is known to exhibit values larger than 
one~\cite{Levkivskyi2008}, for our setup this quantity diverges for values of bias $V$ such that the direct contributions to the 
differential conductance vanishes. The above peculiar  behavior  can be harnessed to improve the performances of the MZI. For example, via  Eq.~(\ref{Eq:resonance}), the bias $V$ can be used as a knob for fine tuning of the resonance condition, as the precise value of the momentum difference $\Delta k$ is a priori unknown. We finally note that for systems which do not feature a modulation of the tunneling ($\lambda \rightarrow \infty$), Eq.~(\ref{Eq:resonance}) predicts  that charge transfer can be achieved for sufficiently high bias, as an interaction-driven resonance is met for a threshold voltage $eV_{\rm th}=v_F\Delta k$. This picture shares analogies with the experimental findings of Deviatov {\it et al.} ~\cite{Deviatov}, who reported inter-edge transport and AB oscillations only beyond  a threshold bias.

{\it Conclusions. --- } We have  shown that  
e-e interactions reduce the performances of  a MZI with co-propagating edge states.
There are, however, striking differences with respect to non-simply connected MZI architectures. Indeed, 
while in the latter case the interferometer edge channels are coupled to additional modes that carry information 
away from the system \cite{Neder2006,Roulleau2007,Litvin2008,Levkivskyi2008}, in our setup they only interact 
among themselves  and the information is redistributed in the system. The major impact  is the spoiling of the 
resonant  tunneling condition that realizes the BS (the oscillating component of the interferometer current being 
only marginally affected because inter-edge interactions are negligible in the interference region). This leads to the unexpected result that strong interactions yield a reduction of the current visibility for small voltages, but an 
enhancement for larger voltages, with respect to the non-interacting case. Furthermore, the differential 
conductance can become negative in some voltage range, while its visibility can take large values or even diverge.

{\it Acknowledgments. ---} We thank B. Karmakar for useful discussions. This work has been supported by MIUR through FIRBIDEAS Project No. RBID08B3FM, by EU Project IP-SIQS, by the EU FP7 Programme 
under Grant Agreement No. 234970-NANOCTM, No. 248629-SOLID, No. 233992-QNEMS, No. 238345-GEOMDISS, and 
No. 215368-SEMISPINNET. L. C. acknowledges support from ERC Advanced Grant No. 290846.

\end{document}